\documentclass[superscriptaddress,twocolumn,aps,prb,floatfix,showpacs,amsmath,amssymb]{revtex4}
\usepackage{graphicx}
\usepackage{amsmath,amssymb,amsfonts}
\usepackage{dcolumn}
\usepackage{bm}

\begin{document}
\title{Large Positive Magnetoresistance
of the Lightly Doped La$_{2}$CuO$_4$ Mott Insulator
}
\author{I. Rai\v{c}evi\'{c}}
  \affiliation{National High Magnetic Field Laboratory and Department of Physics, Florida State University, Tallahassee, FL 32310, USA}
\author{Dragana Popovi\'c}
\affiliation{National High Magnetic Field Laboratory and Department of Physics, Florida State University, Tallahassee, FL 32310, USA}
\author{C. Panagopoulos}
\affiliation{Department of Physics, University of Crete and FORTH,
71003 Heraklion, Greece} \affiliation{Division of Physics and
Applied Physics, Nanyang Technological University, Singapore}
\author{T. Sasagawa}
\affiliation{Materials and Structures Laboratory, Tokyo Institute
of Technology, Kanagawa 226-8503, Japan}
\date{\today}

\begin{abstract}
The in-plane and out-of-plane magnetoresistance (MR) of single crystals of La$_{2}$CuO$_4$, lightly doped ($x=0.03$) with either Sr (La$_{2-x}$Sr$_x$CuO$_4$) or Li (La$_2$Cu$_{1-x}$Li$_x$O$_4$), have been measured in the fields applied parallel and perpendicular to the CuO$_2$ planes.  Both La$_{1.97}$Sr$_{0.03}$CuO$_{4}$ and La$_{2}$Cu$_{0.97}$Li$_{0.03}$O$_{4}$ exhibit the emergence of a positive MR at temperatures ($T$) well below the spin glass (SG) transition temperature $T_{sg}$, where charge dynamics is also glassy.  This positive MR grows as $T\rightarrow 0$ and shows hysteresis and memory.  In this regime, the in-plane resistance $R_{ab}(T,B)$ is described by a scaling function, suggesting that short-range Coulomb repulsion between two holes in the same disorder-localized state plays a key role at low $T$.  The results highlight similarities between this magnetic material and a broad class of well-studied, nonmagnetic disordered insulators.

\end{abstract}


\maketitle

\section{Introduction}

\begin{figure}[b]
\includegraphics[width=8.2cm]{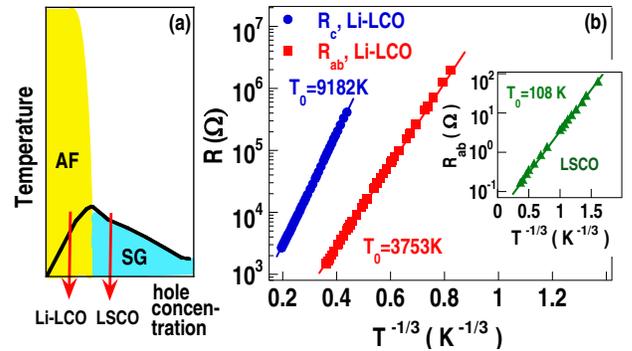}
\caption {(color online)
 (a) Schematic phase diagram of lightly hole-doped cuprates.
The spins are frozen in the region indicated beneath the solid line.  For $x=0.03$, LSCO and Li-LCO are located in different parts of the phase diagram, as shown.  (b)
$R_{c}(T)$ and $R_{ab}(T)$ for LiLCO.  Inset: $R_{ab}(T)$ in LSCO.}\label{fig:RvsT_LiLCO_LSCO}
\end{figure}

The doping of a Mott insulator is a fundamental problem of condensed matter physics that is of relevance to many materials.\cite{Imada-MITreview}  In doped Mott insulators, the presence of several competing ground states combined with a Coulomb repulsion between electrons leads to various nanoscale inhomogeneities.\cite{Elbio} Many different arrangements of these nanoscopic ordered regions often have comparable energies, resulting in slow dynamics typical of glassy systems.  In cuprates, for example, spin glass behavior is well established at $T< T_{sg}(x)$ ($x$ -- doping)\cite{SG1,SG2,SG3,SG4,SG5} [Fig.~\ref{fig:RvsT_LiLCO_LSCO}(a)].  Moreover, in La$_{1.97}$Sr$_{0.03}$CuO$_{4}$ at $T\ll T_{sg}$,
charge heterogeneities are also dynamic, consistent with an underlying charge cluster glass ground state that results from Coulomb interactions.\cite{raicevic08,Glenton}  Even though such glassy freezing of charges may be crucial for the understanding of the transition from an insulator into a conductor in many materials,\cite{Miranda} including cuprates,\cite{Christos-Vlad} studies in the regime of charge glassiness are relatively scarce.

In La$_{2-x}$Sr$_x$CuO$_4$ (LSCO) with $x=0.03$ [Fig.~\ref{fig:RvsT_LiLCO_LSCO}(a)], resistance
noise and impedance spectroscopy at $T\ll T_{sg}$ have revealed\cite{raicevic08,Glenton} that the charge dynamics becomes increasingly slow and correlated, \textit{i.e.} glassy, as $T\rightarrow 0$.
In the same $T$ regime, the out-of-plane or $c$-axis resistance $R_c$ in $B\parallel c$ showed signatures of glassiness, such as hysteresis and memory \cite{raicevic08}.  These results are consistent with the picture of collective charge rearrangements in the hole-rich regions in the presence of the hole-poor antiferromagnetic (AF) domains in CuO$_2$ planes.  Each AF domain is known to have a weak ferromagnetic (FM) moment associated with it, such that the direction of the FM moment is uniquely linked to the phase of the AF order.\cite{thio88,thio90,lavrovSS01}  Although MR $R_c(B\parallel c)$ at such low $T$ was positive,\cite{raicevic08} in contrast to most MR reports on LSCO and other cuprates, it was not studied in detail.  Moreover, the in-plane transport, which is generally agreed to be more relevant to the physics of cuprates, was not measured.  In the present work, in an effort to understand the origin of the peculiar hysteretic positive $R_c(B\parallel c)$ and to elucidate the low-$T$ properties of this system, measurements of both $R_{c}(B,T)$ and in-plane $R_{ab}(B,T)$ in LSCO with $x=0.03$ have been performed over a wide range of $T$ and $B$.
%
%
\begin{figure*}[t]
\begin{minipage}[c]{0.24\textwidth}
\includegraphics[width=4.35cm,height=4.8cm]{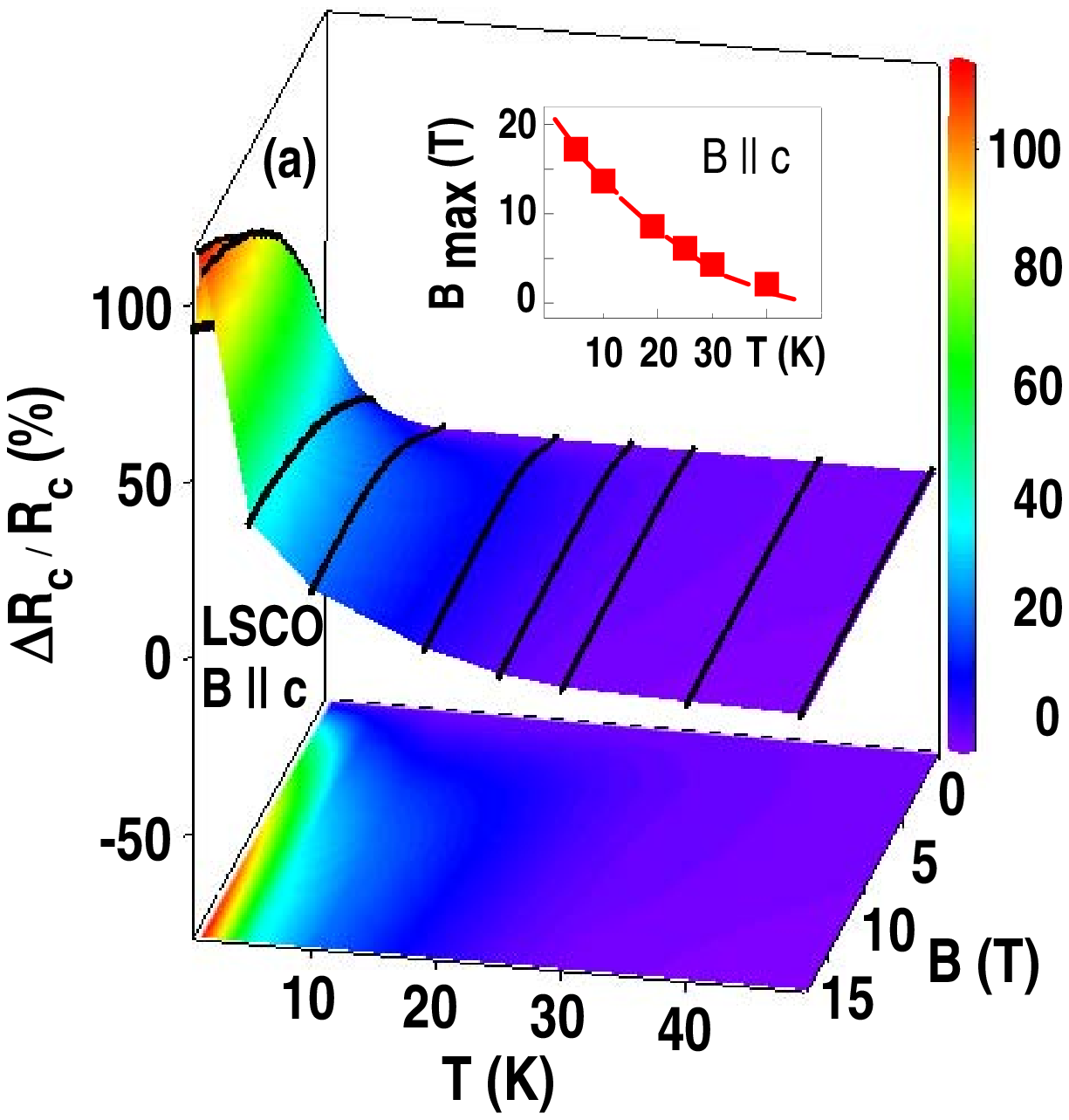}
\end{minipage}
\begin{minipage}[c]{0.24\textwidth}\includegraphics[width=4.35cm,height=4.8cm]{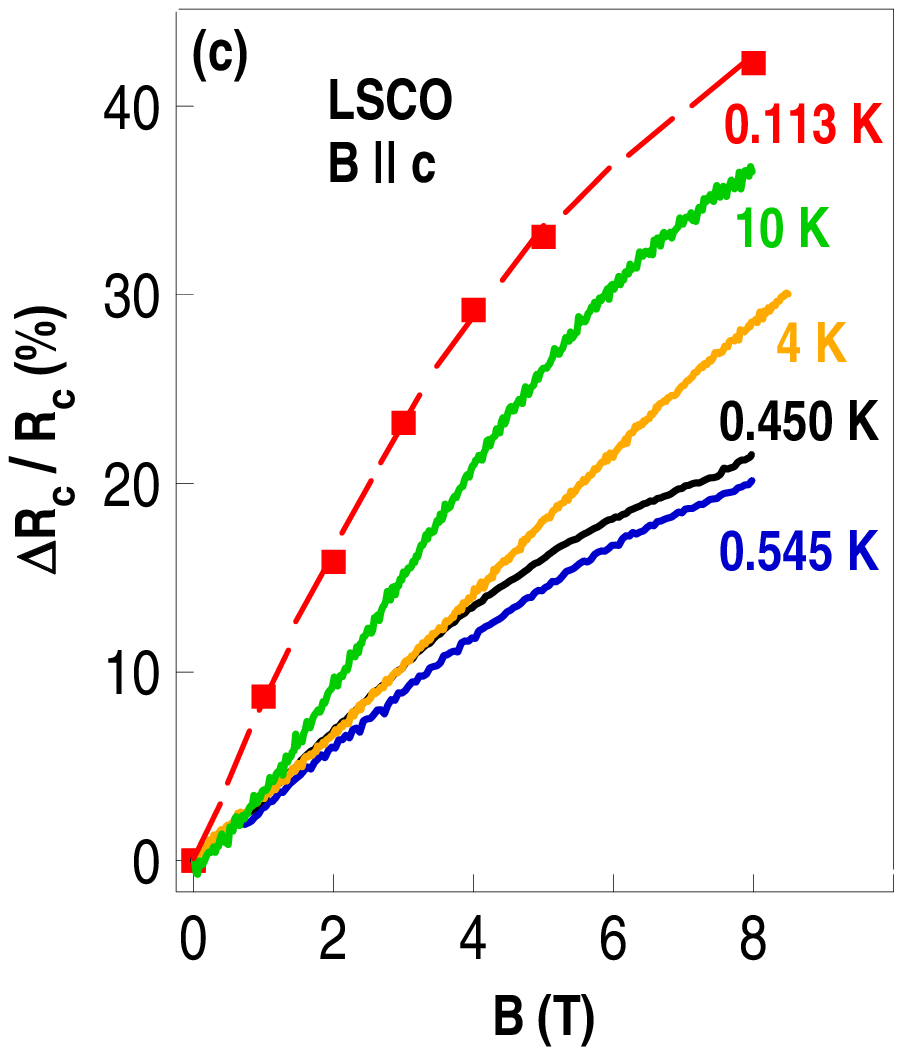}
\end{minipage}
\begin{minipage}[c]{0.24\textwidth}\includegraphics[width=4.35cm,height=4.8cm]{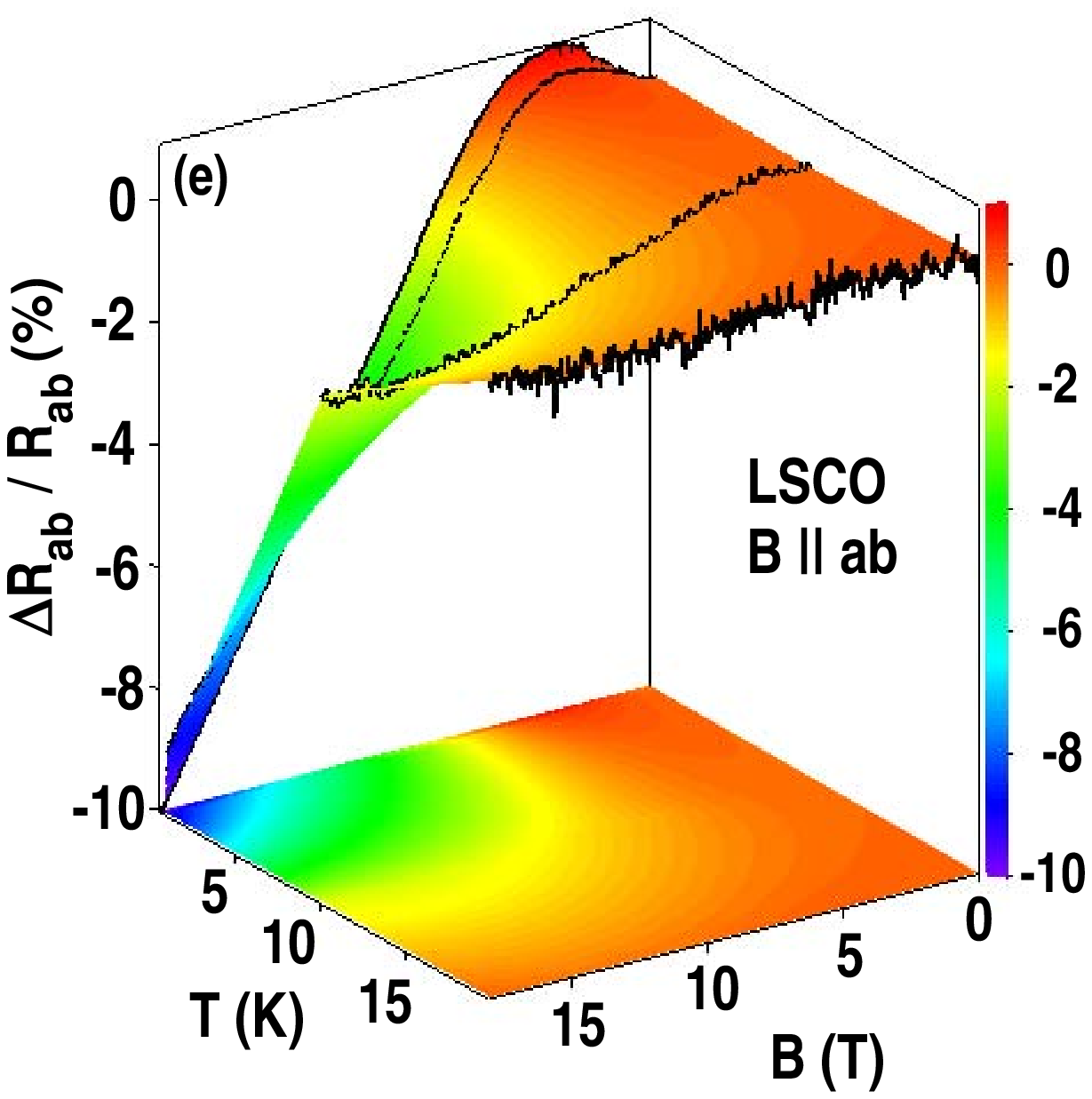}
\end{minipage}
\begin{minipage}[c]{0.24\textwidth}\includegraphics[width=4.35cm,height=4.8cm]{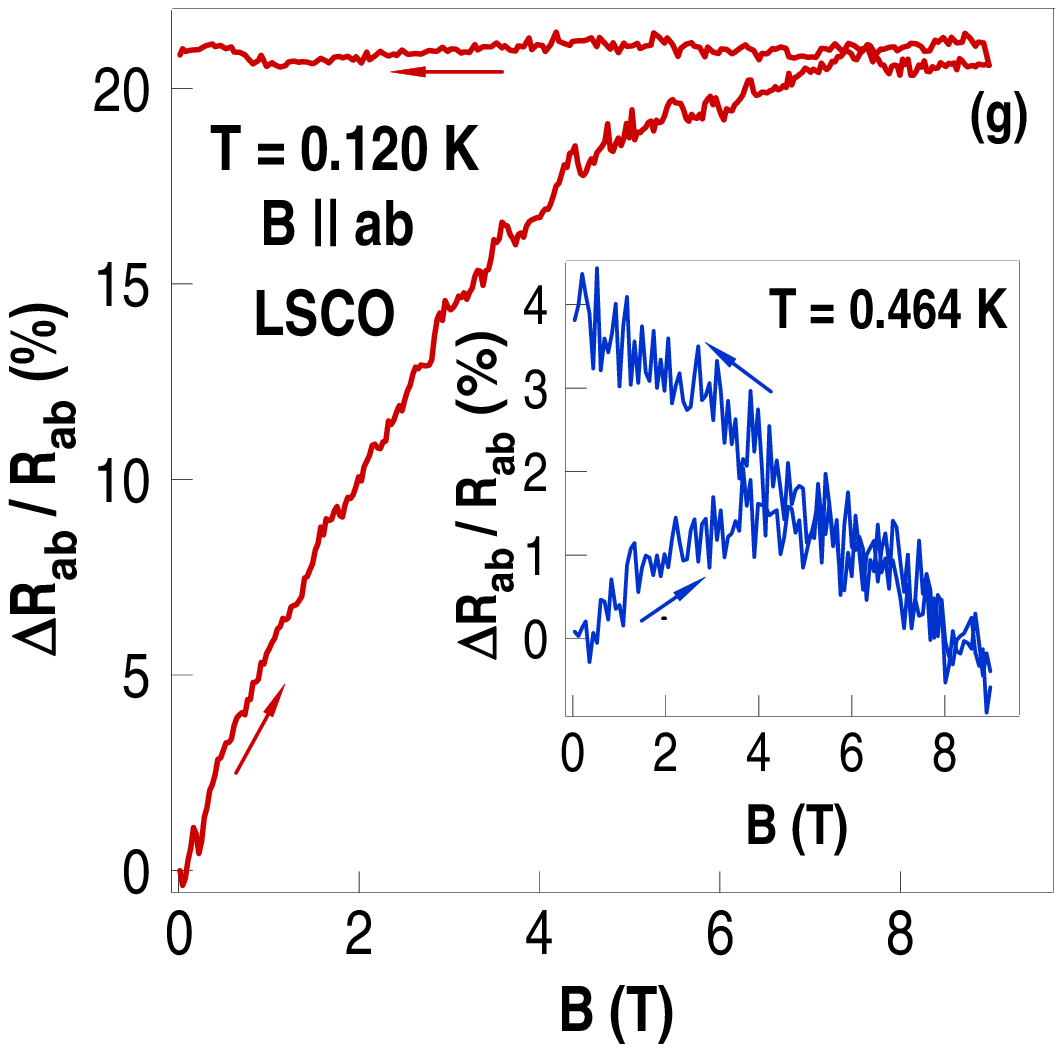}
\end{minipage}
\begin{minipage}[c]{0.24\textwidth}\includegraphics[width=4.35cm,height=4.8cm]{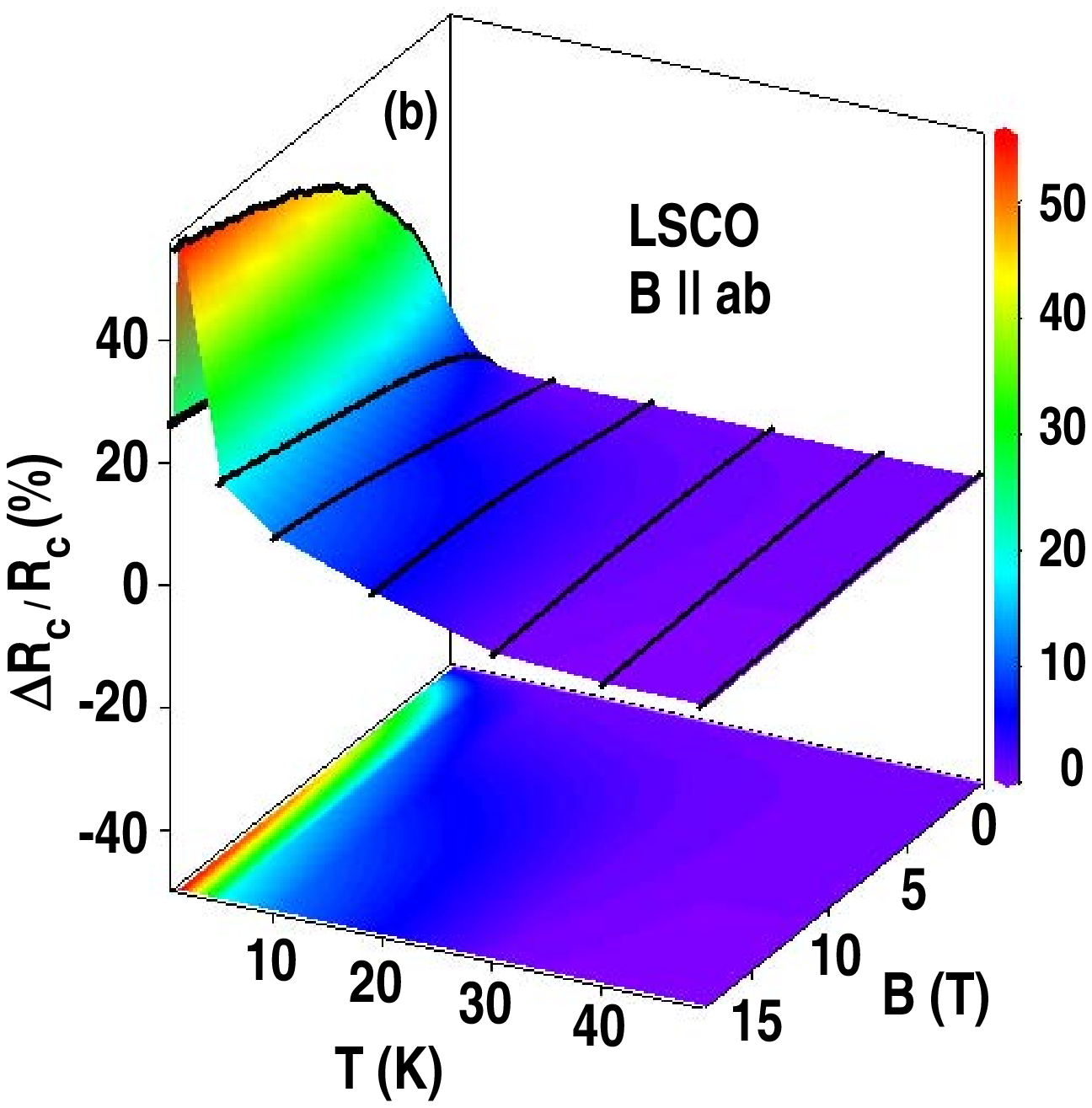}
\end{minipage}
\begin{minipage}[c]{0.24\textwidth}\includegraphics[width=4.35cm,height=4.8cm]{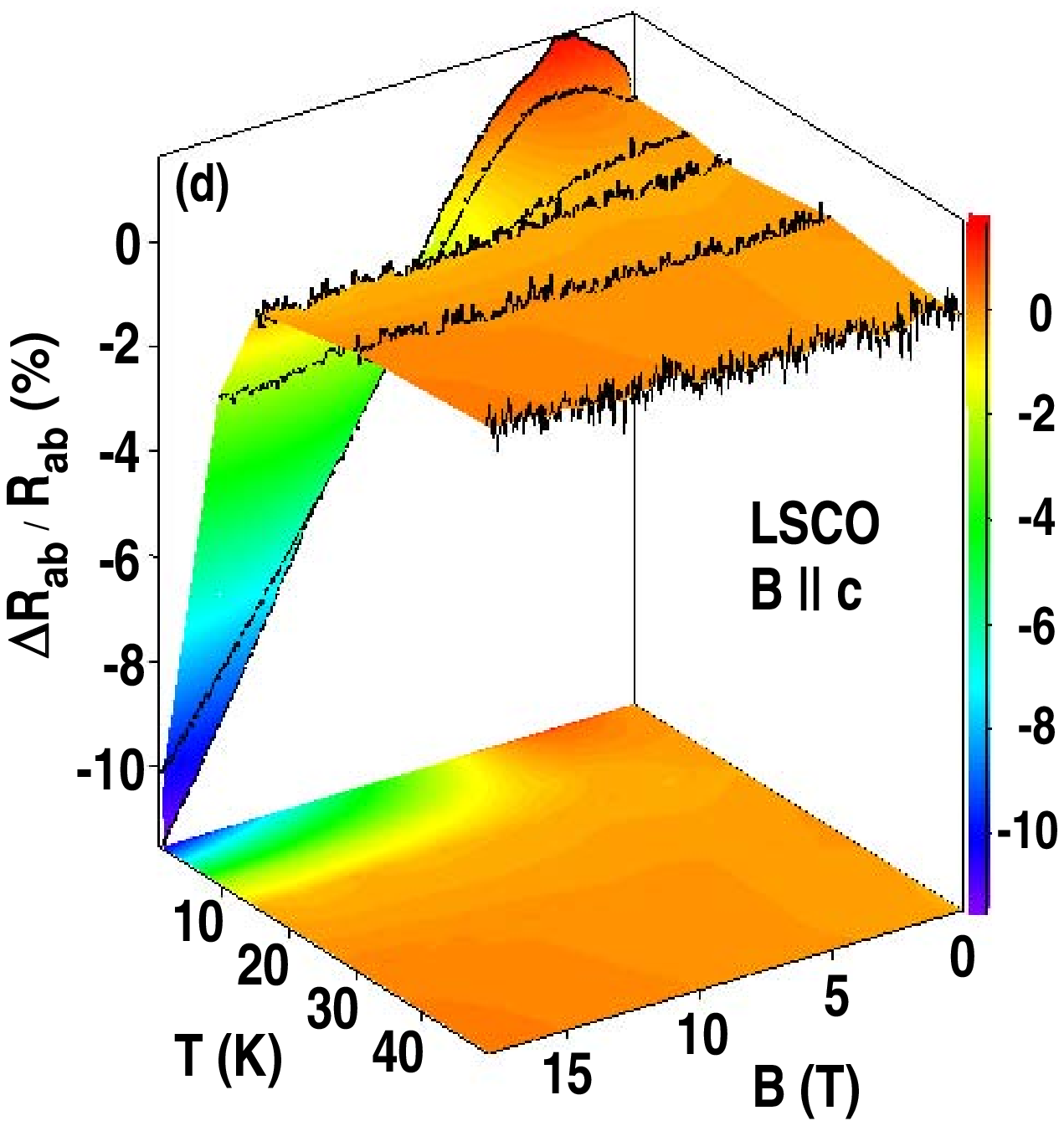}
\end{minipage}
\begin{minipage}[c]{0.24\textwidth} \includegraphics[width=4.35cm,height=4.8cm]{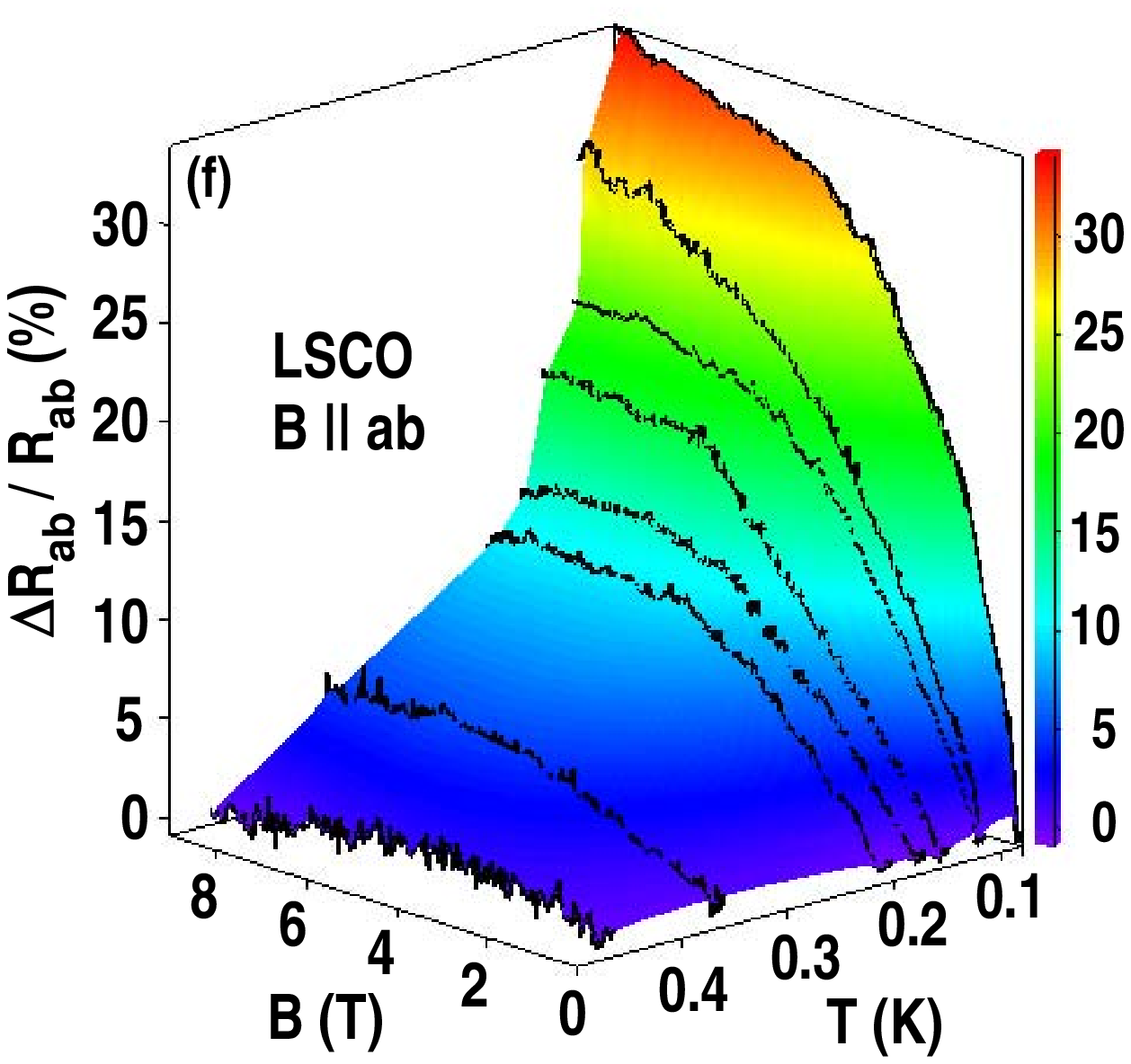}
\end{minipage}
\begin{minipage}[c]{0.24\textwidth} \includegraphics[width=4.35cm,height=4.8cm]{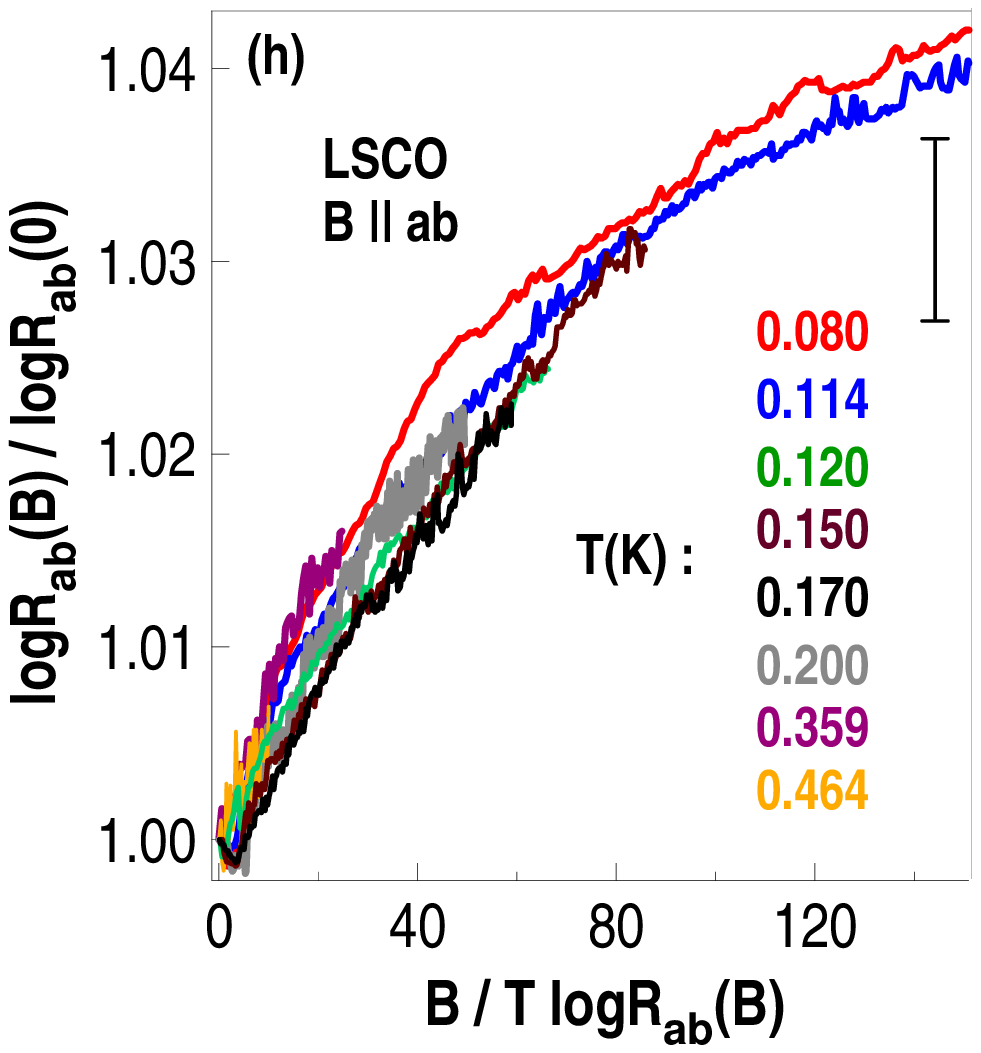}
\end{minipage}

\caption{(color online) $c$-axis MR (sample 2) for (a)
$B\parallel c$ and (b) $B\parallel ab$, at $T\geq 0.6$~K. Inset in (a): The value of the field at the $R_{c}(B)$ maximum for different $T$; the dashed line is a phenomenological fit $B_{max}$[T]$=22.8\exp(-T$[K]$/16.2)$.  (c) Low-$T$ out-of-plane MR (sample 1) for $B\parallel c$ shows nonmonotonic dependence on $T$ below $\sim 1$~K. In-plane MR for (d) $B\parallel c$ and (e) $B\parallel ab$, at $T\geq 0.6$~K. (f) In-plane MR for $B\parallel ab$ at $T<0.5$~K.  (g) Hysteretic behavior of the positive component of the in-plane MR for $B\parallel ab$
at $T=0.120$~K and $T=0.464$~K (inset). The arrows denote the direction of
$B$-sweeps.  (h) The scaling of the $R_{ab}$ data, shown in (f), with $T$ and $B$.  The error bar corresponds to the maximum $R_{ab}$ change due to $T$ fluctuations.
}\label{fig:LSCOMR}
\end{figure*}
%
Furthermore, in order to help disentangle the effects of interactions, disorder, and magnetism, studies were also extended to La$_{2}$Cu$_{1-x}$Li$_{x}$O$_{4}$ (Li-LCO) with $x=0.03$.

LSCO and Li-LCO have the same parent compound, La$_{2}$CuO$_4$.  While Sr dopants are located out-of-plane, Li dopants replace Cu directly in plane.  This substitution removes a Cu$^{2+}$ spin, since Li$^+$ does not have any magnetic moment. In addition, just like Sr, each Li dopant introduces one hole into the $ab$ plane.  The charge carriers frustrate the magnetism, leading to the destruction of the long-range AF order at $x=0.02$ in LSCO\cite{kastner98} and $x\gtrsim 0.03$ in Li-LCO.\cite{sasagawa02}  In the $x=0.03$ Li-LCO studied here, the long-range AF order is still present in the experimental $T$-range [Fig.~\ref{fig:RvsT_LiLCO_LSCO}(a)].\cite{Neel-comment}  In spite of the differences in the type and strength of the disorder, the structural\cite{Sarrao96} and magnetic\cite{sasagawa02} properties of LSCO and Li-LCO are nearly identical at low $x$, including the emergence of the spin glass phase.
Furthermore, dielectric response provides evidence for slow \cite{park05,Glenton} and glassy \cite{park05} charge dynamics in Li-LCO at low $x$.
However, Li-LCO remains an insulator at all dopings,\cite{kastner88} whereas LSCO is a high-temperature superconductor (HTS) for $0.055\leq x <0.27$.  Therefore, a detailed comparison of Li-LCO and LSCO is valuable for the understanding of the low-$T$ charge transport properties of lightly doped cuprates on the border of magnetism, and in the case of LSCO, in the pseudogap regime, near the border with unconventional superconductivity.

Here we focus on the previously unexplored magnetoresistance at $T\ll T_{sg}$ in La$_2$CuO$_4$, lightly doped with either Sr or Li, where charge dynamics is glassy.  Unexpectedly, in both materials we observe the emergence of a strong, hysteretic, \textit{positive} MR in both in-plane and out-of-plane transport, regardless of the direction of $B$.  Surprisingly, in spite of the presence of the
AF order (long-range in La$_{2}$Cu$_{0.97}$Li$_{0.03}$O$_{4}$ and short-range in La$_{1.97}$Sr$_{0.03}$CuO$_{4}$), the lightly doped La$_2$CuO$_4$ behaves essentially the same as various nonmagnetic, disordered materials with strong Coulomb interactions.

\begin{figure*}[t]
\begin{minipage}[c]{0.32\textwidth}
\includegraphics[width=5.8cm,height=4.8cm]{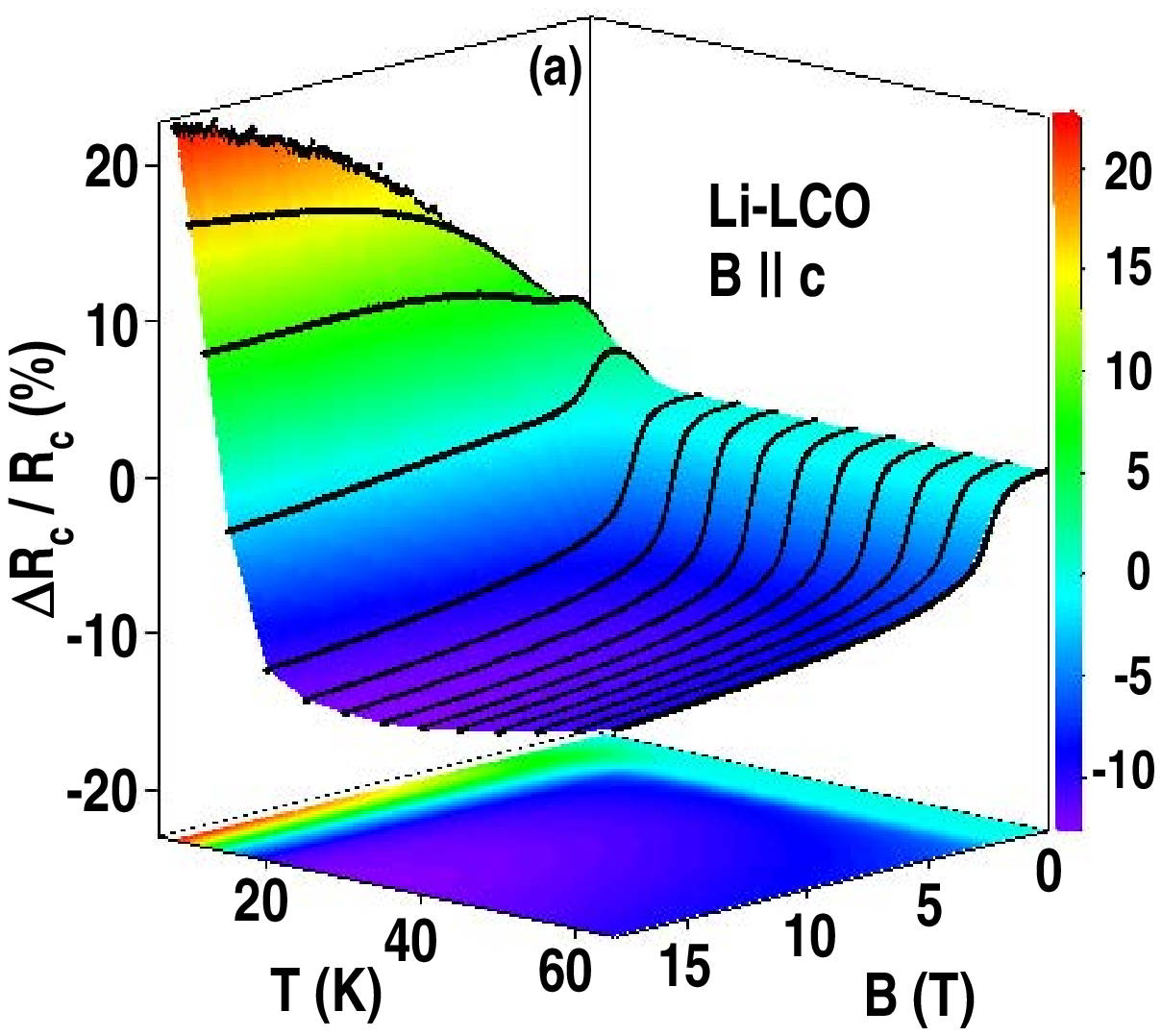}
\end{minipage}
\begin{minipage}[c]{0.32\textwidth} \includegraphics[width=5.8cm,height=4.8cm]{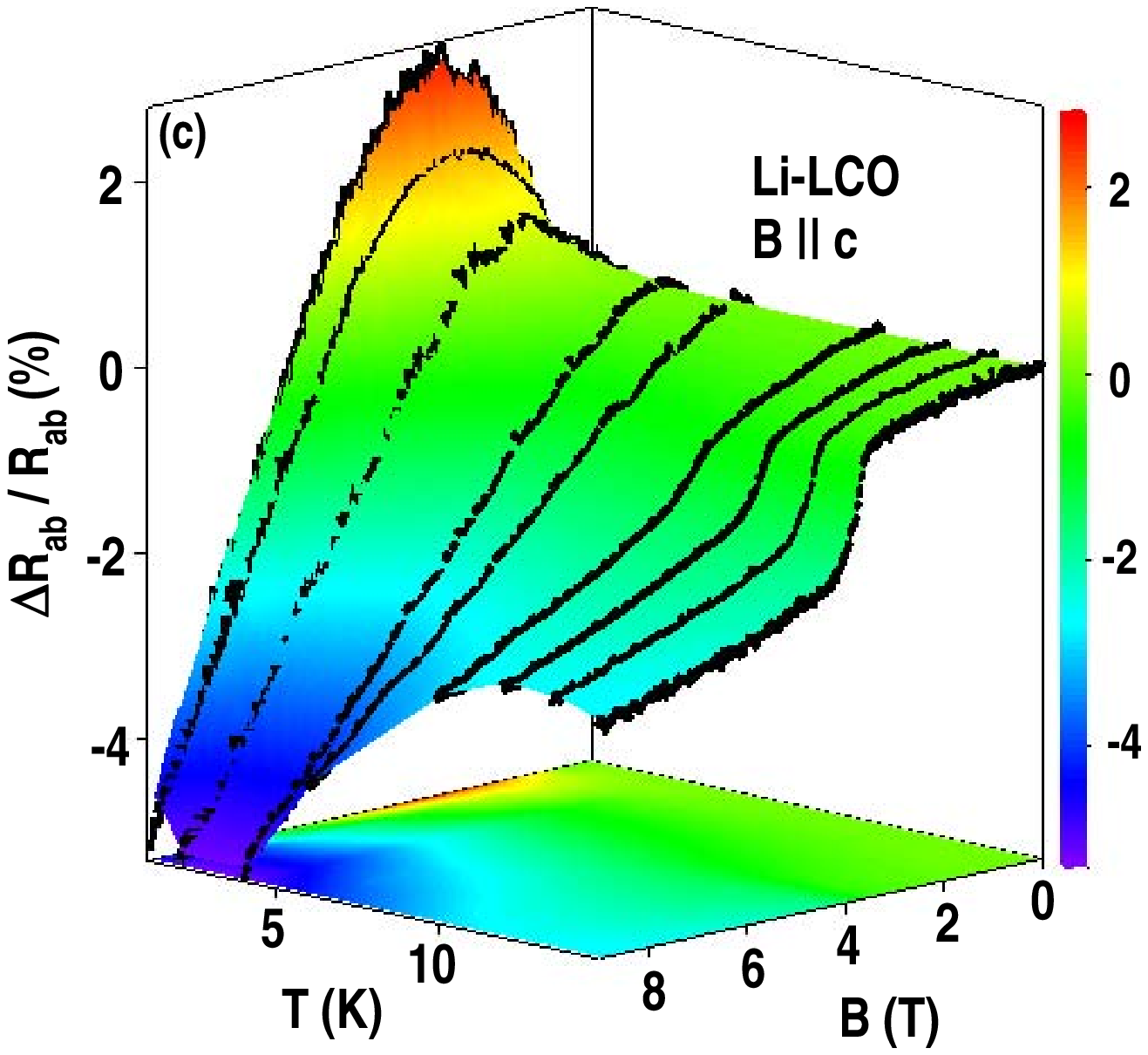}
\end{minipage}
\begin{minipage}[c]{0.32\textwidth} \includegraphics[width=5.8cm,height=4.8cm]{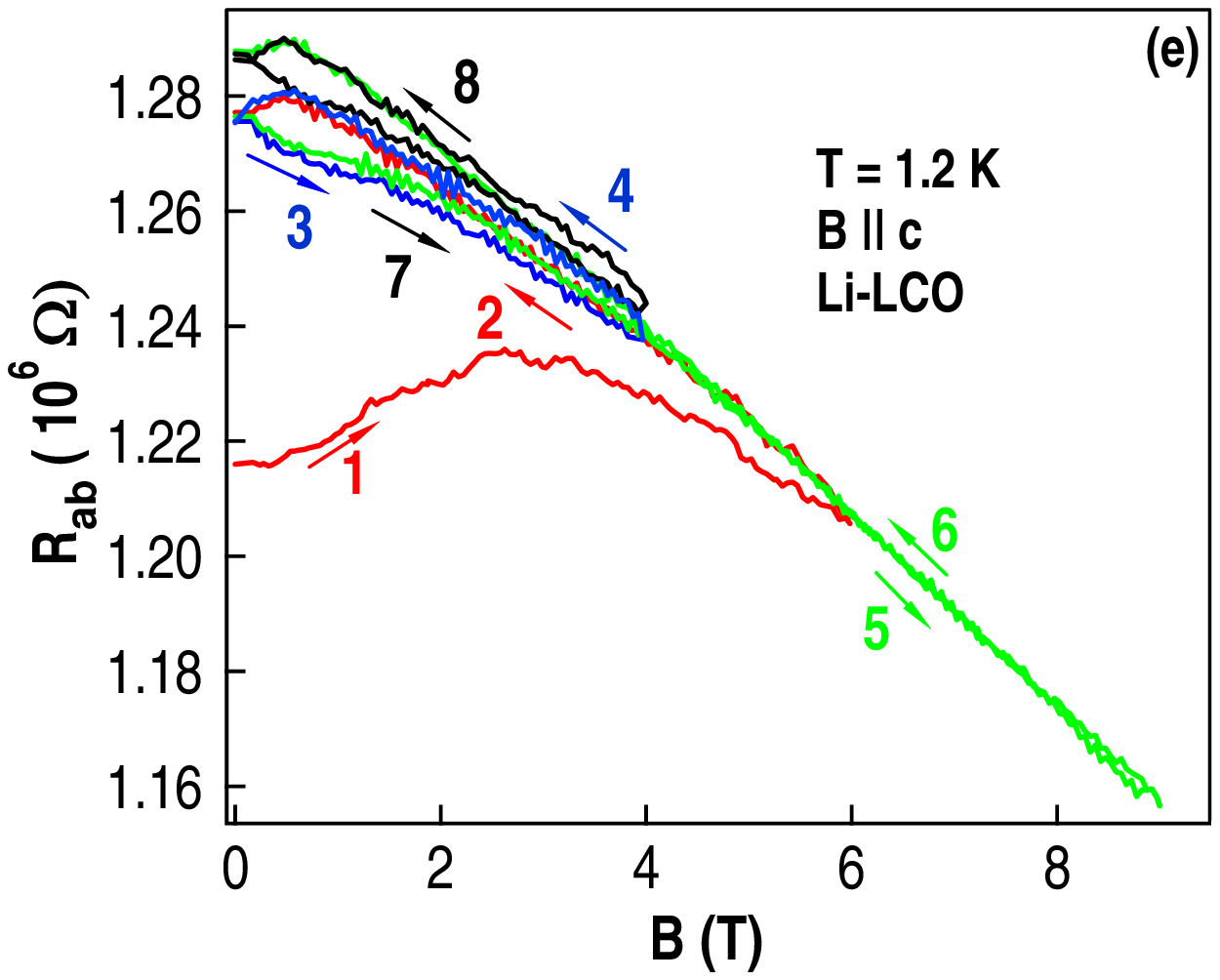}
\end{minipage}
\begin{minipage}[c]{0.32\textwidth} \includegraphics[width=5.8cm,height=4.8cm]{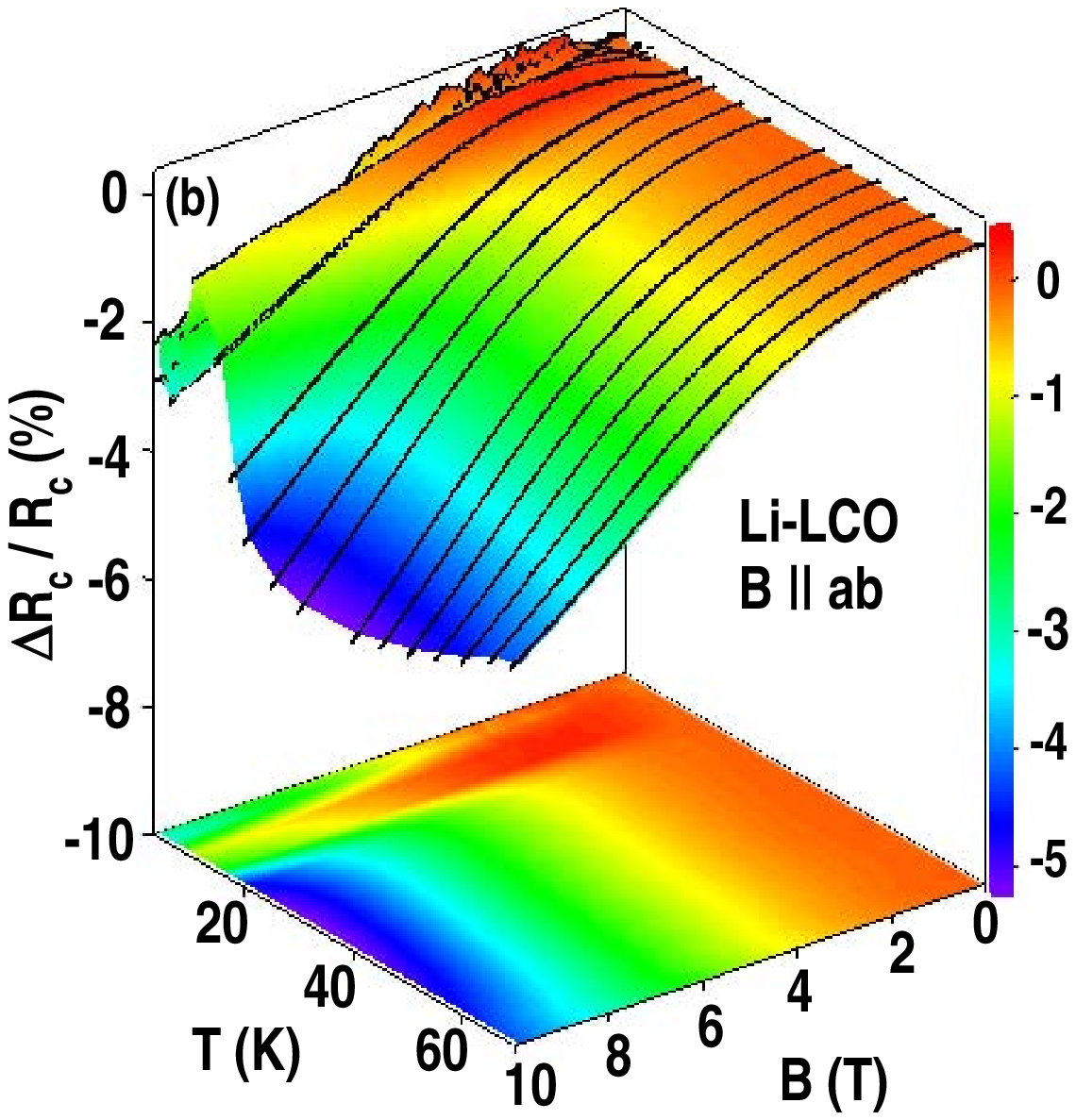}
\end{minipage}
\begin{minipage}[c]{0.32\textwidth} \includegraphics[width=5.8cm,height=4.8cm]{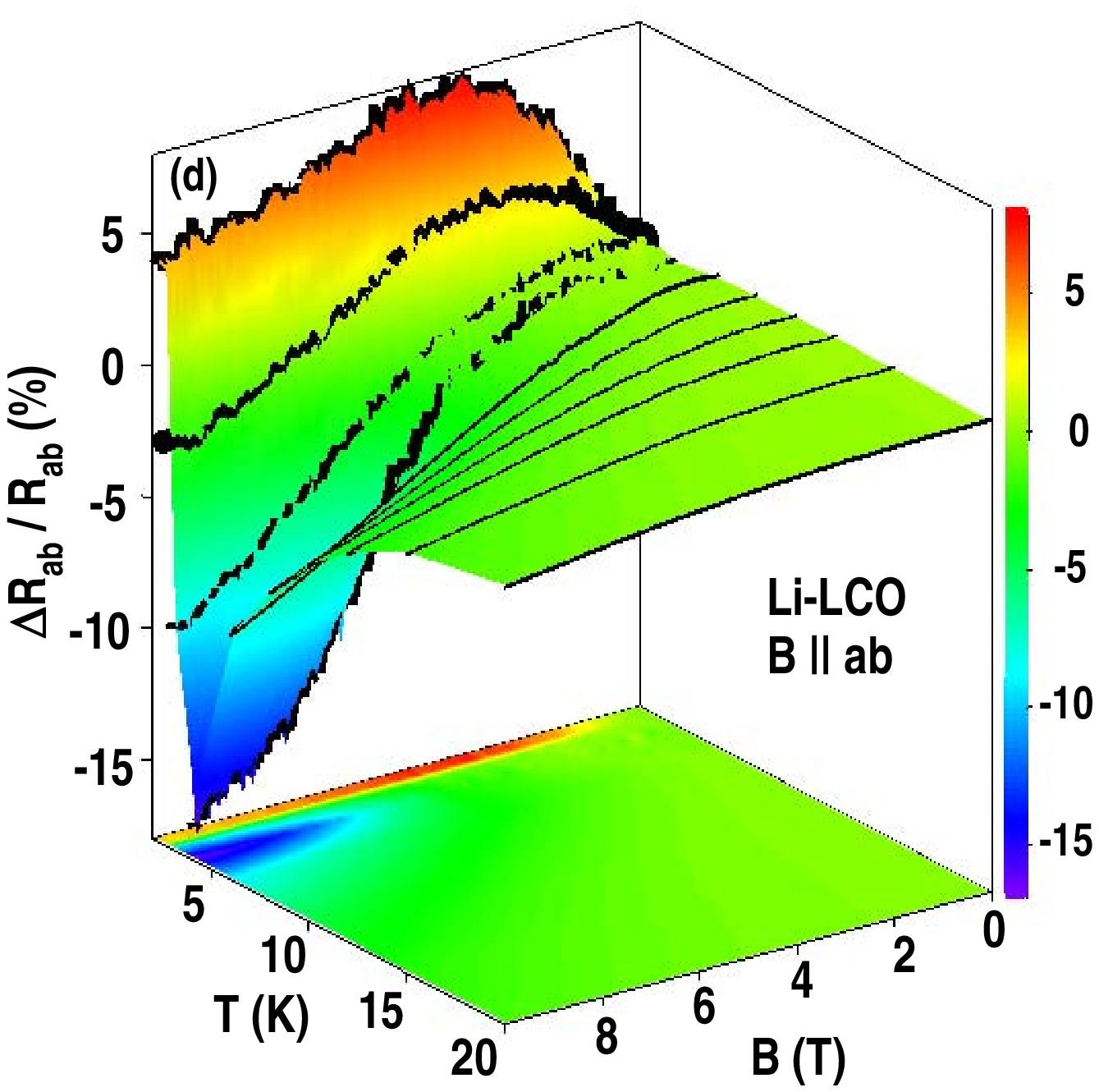}
\end{minipage}
\begin{minipage}[c]{0.32\textwidth} \includegraphics[width=5.8cm,height=4.8cm]{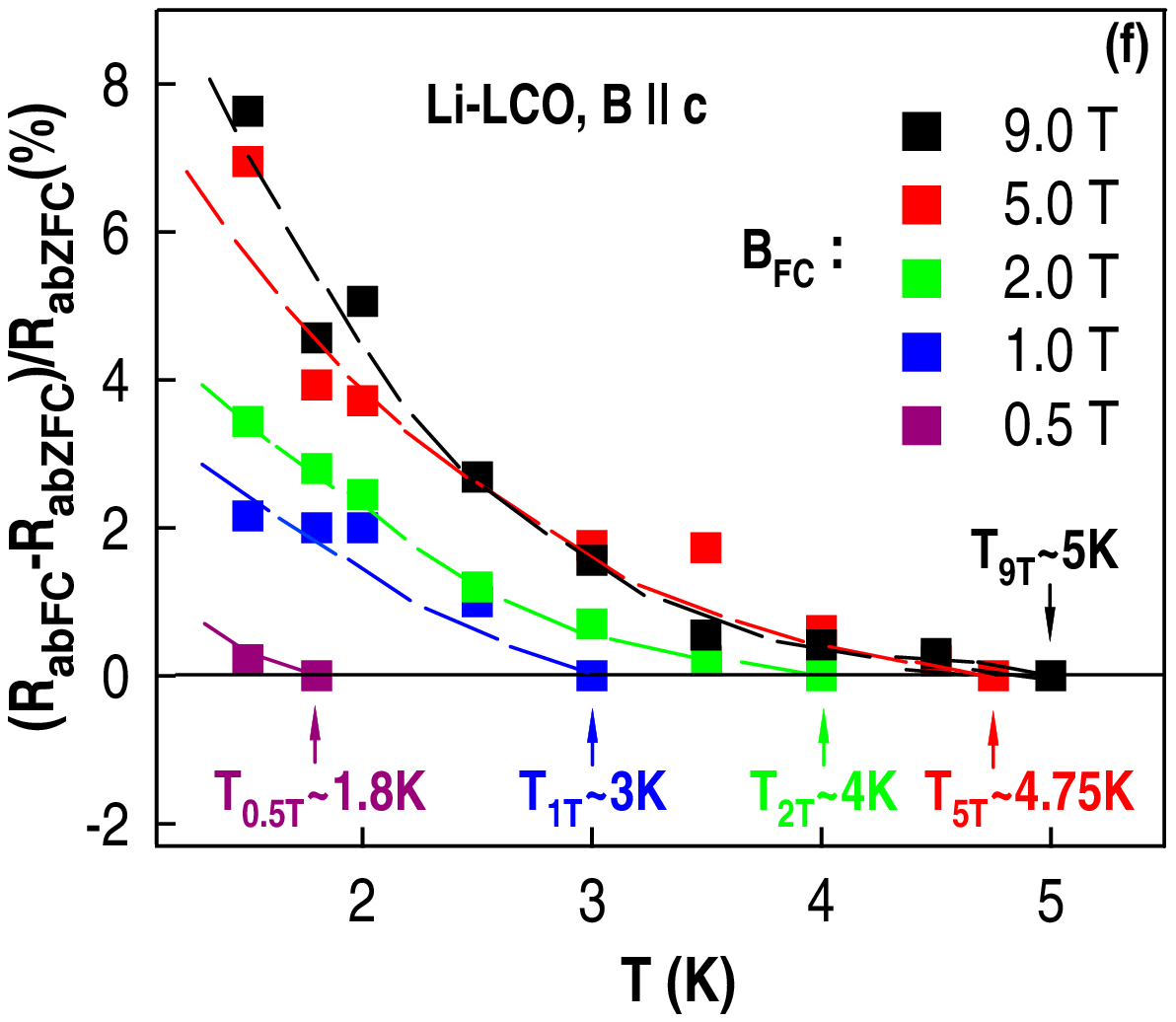}
\end{minipage}

\caption{(color online) $c$-axis MR for (a)
$B\parallel c$ (LMR; $T\geq 6$~K) and (b) $B\parallel ab$ (TMR; $T\geq 3$~K).  In-plane MR for (c) $B\parallel c$ (TMR; $T\geq 1$~K) and (d) $B\parallel ab$ (LMR; $T\geq 2$~K).  (e) The hysteresis in $R_{ab}(B)$. The arrows and the
numbers denote the direction and the order of $B$-sweeps.  The two subloops are incongruent, indicating that domains interact.  (f) The difference between FC and ZFC $R_{ab}$ \textit{vs.} $T$ for different fields $B_{FC}$ applied during cooling, as shown. Dashed lines guide the eye.}\label{fig:LiLCOMRHYSTERESIS}
\end{figure*}


\section{Experiment}

Single crystals of LSCO and Li-LCO with a nominal $x=0.03$ were grown by the traveling-solvent floating-zone technique.\cite{sasagawa98, sasagawa02}  For both LSCO\cite{raicevic08} and Li-LCO\cite{sasagawa02} samples, spin freezing occurs at $T_{sg}\sim 7-8$~K.  Five samples were cut out and polished into bars with the following dimensions: $2.10 \times 0.44 \times 0.42$ mm$^{3}$ (LSCO) and $2.2 \times 0.57 \times 0.41$ mm$^{3}$ (Li-LCO) for $R_{ab}$, $0.6 \times 0.8 \times 1.57$ mm$^{3}$ [LSCO, sample 1 (Ref. \onlinecite{raicevic08})],
$0.6 \times 0.9 \times 1.6$ mm$^{3}$ [LSCO, sample 2 (Ref. \onlinecite{raicevic08})] and $0.5 \times 0.44 \times 1.6$ mm$^{3}$ (Li-LCO) for $R_c$ measurements.  $R$ was measured with a standard 4-probe ac method ($\sim7$~Hz) in the Ohmic regime.  For $R_{ab}$ measurements, the current $I\parallel B$.  $B$ was swept at a fixed $T$, with sweep rates low enough (0.001-0.3 T/min) to avoid the sample heating.

\section{Results}

In both materials, $R_{ab}$ and $R_c$ exhibit insulating behavior in the entire experimental $T$ range.  In particular, in the regime of interest, where novel positive MR emerges at low $T$, all resistances obey the variable-range hopping (VRH) law $R=R_{0}\exp(T_{0}/T)^{1/3}$ (Fig.~\ref{fig:RvsT_LiLCO_LSCO}(b), Ref. \onlinecite{raicevic08}).

In LSCO, the $c$-axis MR in both longitudinal (LMR; $B\parallel c$) and transverse (TMR; $B\parallel ab$) configurations (Figs.~\ref{fig:LSCOMR}(a) and ~\ref{fig:LSCOMR}(b), respectively) are very weak and negative at the highest $T=50$~K.  However, below 40~K, a positive MR (pMR) appears at low $B$ and becomes very large ($\sim 50-100$\%) at low $T\sim 1.5$~K.  The maximum in the MR curves shifts to increasingly high fields as $T$ is lowered (Fig.~\ref{fig:LSCOMR}(a) inset).  We note that the LMR is always significantly larger than the TMR, similar to the observations at higher $x$.\cite{Hussey}  In analogy with the AF YBa$_2$Cu$_3$O$_{6+x}$ (YBCO) study,\cite{andoYBCO01} the positive $c$-axis MR may be understood to result from the field suppression of the spin fluctuations, leading to an increased hole confinement\cite{andoAF01} and thus reduced hopping between the
hole-rich regions in neighboring CuO$_2$ planes.\cite{andocaxis02}

The monotonic increase in the magnitude of the pMR with decreasing $T$ stops at $T\sim$1.5 K for both $B$
orientations, and the MR begins to drop [Figs.~\ref{fig:LSCOMR}(a) and ~\ref{fig:LSCOMR}(b)].  A closer inspection of the low-$T$ MR on another sample shows [Fig.~\ref{fig:LSCOMR}(c)] that this initial drop of the MR is reversed below $\sim $1~K.  In that regime, the pMR is increasingly strong again as $T$ is reduced, and the hysteretic and memory effects appear.\cite{raicevic08}  The data suggest that the pMR mechanism changes below $\sim 1$~K, and that it is closely related to the onset of glassiness in transport.\cite{raicevic08}

In contrast to $R_{c}(B)$, $R_{ab}$ for both $B$ orientations is negligible from 50~K down to 10~K, when it becomes negative and increases in magnitude as $T$ is reduced [Figs.~\ref{fig:LSCOMR}(d) and ~\ref{fig:LSCOMR}(e)].
The growth of the negative in-plane MR, clearly related to the spin-glass order (here $T_{sg}\sim 7-8$~K), has been observed before on both LSCO single crystals \cite{andoMR04} and thin films \cite{cieplak04} with the same $x$ as our samples, and attributed to the reorientation of the weak FM moments.\cite{andoMR04} However, as $T$ is reduced further, below 1~K, the low-$B$ pMR emerges again [Figs.~\ref{fig:LSCOMR}(d) and ~\ref{fig:LSCOMR}(e)].  This positive contribution grows rapidly with decreasing $T$, and dominates the MR in the entire experimental $B$-range at the lowest $T\sim 0.1$~K [Fig.~\ref{fig:LSCOMR}(f)].  Surprisingly, only this positive component of the MR shows glassy effects such as hysteresis, memory and magnetic history dependence, as illustrated in Fig.~\ref{fig:LSCOMR}(g).  Thus both $R_{c}$ and $R_{ab}$ exhibit the emergence of the low-$B$ pMR that appears to be related to charge glassiness.

As expected, the behavior of La$_{2}$Cu$_{0.97}$Li$_{0.03}$O$_{4}$ at high $T$ is quite similar to that of the AF ($x=0.01$) LSCO.\cite{andoMR03}  For example, the out-of-plane LMR and TMR (Figs.~\ref{fig:LiLCOMRHYSTERESIS}(a) and
~\ref{fig:LiLCOMRHYSTERESIS}(b), respectively) are both negative at high $T$, the LMR exhibits a steplike decrease caused by the spin-flop transition in every other CuO$_{2}$ plane, and the TMR follows a $\propto B^{2}$
dependence due to a smooth rotation of weak FM moments toward the $B$ direction.  Qualitatively similar behavior is observed in the in-plane MR for both $B$ orientations [Figs.~\ref{fig:LiLCOMRHYSTERESIS}(c) and ~\ref{fig:LiLCOMRHYSTERESIS}(d)].  In analogy to the AF LSCO,\cite{andoMR03} here the negative MR probably also results from an increase of the hole localization length due to the reorientation of the FM moments in CuO$_2$ planes.\cite{kotov07}

Just like in LSCO (Fig.~\ref{fig:LSCOMR}), however, we find that the low-$B$ pMR emerges at low $T$ in Li-LCO in both $R_{c}$ and $R_{ab}$ for both $B$ orientations [Figs.~\ref{fig:LiLCOMRHYSTERESIS}(a)-(d)].  In particular, for $R_{ab}$, the pMR becomes observable for $T< 4$~K, \textit{i.e.} below the spin freezing temperature $\sim 7-8$~K.\cite{sasagawa02}  It is intriguing to examine whether this pMR reflects the onset of charge glassiness
in the same manner as in LSCO.

As illustrated in Fig.~\ref{fig:LiLCOMRHYSTERESIS}(e) for $R_{ab}(B\parallel c)$, the pMR in Li-LCO shows indeed the history dependent, multibranch behavior with the same characteristics as those first described in LSCO for $R_{c}(B\parallel c)$ (Ref. \onlinecite{raicevic08}).  We stress that the hysteresis occurs only in the $B$ region where the pMR was initially observed (path 1) after zero-field cooling (ZFC).  The cycle 5-6 shows clearly the absence of hysteresis for $B>4$~T, where the MR is negative.  The merging of paths 1 and 5 at $B=6$ T demonstrates that the
system exhibits return-point memory, just like in LSCO.\cite{raicevic08}

Other similarities to LSCO in the regime of charge glassiness include the dependence of $R(T,B=0)$ on the cooling protocol.\cite{raicevic08}  In particular, $R(B=0)$ obtained after cooling in field $B_{FC}$ is higher than the ZFC $R(B=0)$, as shown in Fig.~\ref{fig:LiLCOMRHYSTERESIS}(f) for $R_{ab}$ and $B_{FC}\parallel c$.  This difference decreases with increasing $T$, and vanishes at a $T_B$ that grows with $B_{FC}$, at least for low enough $B_{FC}$.

\section{Discussion}

Hence, in both Li-LCO and LSCO, the novel, low-$B$ pMR strongly correlates with charge glassiness at low $T$.  In order to identify the microscopic origin of this pMR, we note that, since Li-LCO does not superconduct at any $x$,
superconducting fluctuations may be ruled out.  Likewise, orbital effects are not relevant, since $R$ does not exhibit the expected\cite{shklovskii84} exponential enhancement with $B$.  We conclude that the pMR must be a spin related effect.  However, it is known that, in the VRH regime, the effect of $B$ on Cu spins leads to a negative,\cite{kotov07} not positive MR.  The remaining possibility is the coupling of $B$ to the spins of doped holes, which populate localized states within the Mott-Hubbard gap $U$ of the parent compound.  Those states have a predominantly oxygen character.  In strongly disordered materials with Mott VRH, it is indeed known that Zeeman splitting in the presence of a Coulomb repulsion $U^{'}$ between two holes in the same disorder-localized state leads to a pMR at low enough $T\ll U^{'}$ by blocking certain hopping channels.\cite{KK}  Such a MR is described\cite{Meir} by a universal function of $B/T\log_{10}R$.  It has been shown\cite{Meir} that this function provides a good fit to the MR data in systems as diverse as quasi-2D In$_2$O$_{3-x}$ films\cite{Frydman} and in-plane transport in 3D Y$_{1-x}$Pr$_x$Ba$_2$Cu$_3$O$_7$.\cite{Greene-pMR}

In order to test whether the above mechanism can describe also the pMR data in lightly doped La$_2$CuO$_4$, we analyze the in-plane transport in LSCO at the lowest $T<0.5$~K [Fig.~\ref{fig:LSCOMR}(f)].  We find that all the $R_{ab}(T,B)$ data in the regime of pMR do collapse onto one function of a single scaling parameter [Fig.~\ref{fig:LSCOMR}(h)].  The scaling fails at higher $T$, as expected,\cite{Meir} and at higher $B$, where another mechanism leads to a negative MR.\cite{kotov07}  The in-plane pMR in Li-LCO obeys the same scaling,
but the measurements could not be extended to $T<1$~K because of the high sample resistance.  Using the localization length $\xi\sim 90$~\AA\, obtained from the VRH fits\cite{DOS-comment} and assuming the dielectric constant $\varepsilon\sim 100$ (Ref. \onlinecite{kastner98}), we estimate $U^{'}\sim 18$~K in LSCO, so that $T\ll U^{'}$ is indeed satisfied in the experiment.\cite{xi-Li}  Thus, while unimportant at high $T$, the Coulomb repulsion between two holes in the same localized state plays a dominant role in the low-$T$ MR of lightly doped La$_2$CuO$_4$.

We remark that the out-of-plane transport in cuprates has been generally more difficult to understand and, accordingly, the $R_c(B\parallel c)$ curves at the lowest $T$ [Fig.~\ref{fig:LSCOMR}(c)] cannot be collapsed in the same manner.  It is interesting that they do collapse as a function of $\alpha(T)B/T\log_{10}R$ (not shown), where $\alpha(T)$ is an empirically determined parameter.  This suggests the presence of some additional mechanism that may be captured by generalizing the model of Ref.~\onlinecite{Meir}, but we do not wish to speculate further.

Since previous studies on LSCO\cite{andoMR04,cieplak04} reported only an increasingly negative MR as $T$ was reduced below $T_{sg}$ [see also Figs.~\ref{fig:LSCOMR}(d) and ~\ref{fig:LSCOMR}(e)], the dramatic, qualitative change in the behavior of the MR from negative to positive at even lower $T$ is very surprising.  There was no reason to expect the emergence of the low-$T$ pMR in doped La$_2$CuO$_4$ on theoretical grounds either.   On the other hand, similar large pMR\cite{Frydman,Zvi,Mertes,Goldman,Butko} and charge glassiness\cite{Zvi,Goldman,DP_relax-aging} have been observed in various nonmagnetic, disordered materials with strong Coulomb interactions.  The observed scaling of $R_{ab}(T,B)$ thus indicates that, to leading order, the magnetic background remains inactive.  This is consistent with the picture of AF domains, frozen at low $T$, and holes confined to the domain walls.  While low $B$ may be expected to produce some motion of the domain walls leading to a hysteretic MR, the main transport mechanism within the domain walls that gives rise to the pMR should remain unchanged.  Much higher $B$ will lead to the reorientation of the weak FM moments of the AF domains and the associated negative MR,\cite{kotov07} as observed.

The charge glass observed in lightly doped La$_2$CuO$_4$ thus seems analogous to that in other disordered, interacting systems, except that here only holes in the domain walls contribute to transport and glassiness.  The existence of such a charge cluster glass was also inferred from the noise and dielectric studies.\cite{raicevic08,Glenton}  A model of a gapped insulator with a short-range repulsive interaction shows\cite{sudip} that the disorder-induced localized states in the gap near the chemical potential are located inside the domain walls, and they are expected to lead to glassy dynamics, similar to our observations.  While the data suggest that a charge glass ground state might be universal in strongly interacting, disordered systems, the question of how it evolves into a HTS in some cases (\textit{e.g.} LSCO) and not in others (\textit{e.g.} Li-LCO) remains open.  However, the novel hysteretic pMR provides a practical tool for detecting an underlying charge glassiness confined to the domain walls.  This method could be applied at higher $x$ and in systems with less disorder, such as YBCO, to determine whether charge inhomogeneities are intrinsic or driven by disorder.  In fact, there is some preliminary evidence of the hysteretic pMR in the N\'{e}el state in YBCO at low $T$.\cite{Ando-YBCO}

\section{Conclusions}

We have demonstrated that a positive magnetoresistance associated with charge glassiness emerges in lightly doped La$_2$CuO$_4$ deep within the SG phase.  This is observed in both in-plane and out-of-plane transport, regardless of the direction of the magnetic field, type of dopant, or presence of long-range or short-range AF order.  It is striking that, as $T\rightarrow 0$, this material shows behavior that is characteristic of systems that are far from any magnetic ordering.

\section{Acknowledgments}

We thank E. S. Choi for susceptibility measurements, X. Shi for technical help, V. Dobrosavljevi\'c, L. Benfatto, and M. B. Silva Neto for discussions, NSF DMR-0403491 and DMR-0905843, NHMFL via NSF DMR-0654118, MEXT-CT-2006-039047, EURYI, and the National Research Foundation, Singapore for financial support.

\vspace*{6pt}


\begin{thebibliography}{50}
\bibitem{DOS-comment} For the 2D density of states,
we used $N_{2}\approx (0.7-4.5)\times 10^{15}$(cm$^2$\, eV)$^{-1}$ ($N=(2-13)$/(eV\, cell) \cite{DOS}).
\end{thebibliography}
\end{document}